# Efficiency enhancement on the solar steam generation by wick materials with wrapped graphene nanoparticles

Xiaojia Li[1, #], Guangqiao Xu[4, #], Guilong Peng[2, 3], Nuo Yang[2, 3], Wei Yu[4, *], Chengcheng Deng[1, *]

[1] School of Energy and Power Engineering, Huazhong University of Science and Technology, Wuhan 430074, P. R. China

[2] State Key Laboratory of Coal Combustion, Huazhong University of Science and Technology, Wuhan 430074, P. R. China

[3] Nano Interface Center for Energy (NICE), School of Energy and Power Engineering, Huazhong University of Science and Technology, Wuhan 430074, P. R. China

[4] School of Environment and Materials Engineering, College of Engineering, Shanghai Second Polytechnic University, Shanghai 201209, P. R. China

## Abstract

Solar steam generation technology can utilize abundant and renewable solar energy for many applications. In this work, we proposed a solar steam generator using wick material with wrapped graphene nanoparticles, and the energy efficiency can reaches up to 80%. Instead of traditional smearing method, the chemical wrapping method was used to better adhere the graphene nanoparticles on the wick materials. Through the SEM morphological results, the graphene nanoparticles are shown to be evenly wrapped across the fibres of the wick material, which have better dispersity and stability. The evaporation rate, instantaneous energy efficiency and the absorptivity of three wick materials with/without nanoparticles under different conditions were compared and analyzed. Among the three different wick materials, the flannel cloth with dense fine hairs can provide three-dimensional contact area for wrapping graphene nanoparticles and thus contribute to better evaporation. Additionally, the influence of two different reduction methods and different concentrations of graphene oxide solution on the energy efficiency was also discussed. Our work offers a simple and effective way of using nanotechnology in practical application for solar steam generation.

# 1. Introduction

Nowadays fresh water in the world is in a shortage[1], while sea water is abundant. However, sea water can't be used directly in our daily life due to the high salinity. Meanwhile, solar energy is an abundant, renewable and eco-friendly energy. Therefore, using solar energy to generate fresh water from sea water, i.e. solar desalination, is a promising way to meet our fresh water demand [2-4]. Solar steam generation, as a key process in solar desalination, becomes one of the most popular research topic recently. Besides solar desalination, solar steam generation technology also has many other applications, such as water purification[5, 6], power generation[7-9], oil recover [10, 11] and so on.

To improve the solar steam generation efficiency, many works have been done during the past decades[12-17]. The traditional way for steam generation is using a black plate to absorb the solar energy and heat the water on it to evaporate. In order to improve the evaporation efficiency, various types of solar still like the stepped solar still[18] and wick type solar still[16] have been developed. A lot of improvements have also been applied to the solar still, such as using reflector[4], designing double layer structure[19], using film cooling[15] and so on. However, the overall energy efficiency of these solar stills is usually below 50% due to the large heat loss.

Recently, in order to overcome the bottleneck of the evaporation efficiency of traditional technology, some researchers tried to apply the nanotechnology to improve solar steam generation. For example, the efficiency of steam generation can be improved by dispersing some nanoparticles in water, like ZnO[20], $Al_2O_3$[20, 21], CuO[22], Au[23-25] and graphene[26]. The water with dispersed nanoparticles forms nanofluid, which can increases the water temperature and thereby increase the evaporation efficiency quite well[27, 28]. The evaporation efficiency can reach 69% under 10kw/$m^2$ solar intensity[26] and 80% under 220kw/$m^2$ solar intensity[29]. However, the nanoparticles tend to aggregate and deposit when used for solar steam generation, so the stability of nanoparticles remains a tough challenge[30, 31].

In this paper, wick materials with wrapped graphene nanoparticles are applied to a low-cost solar steam generator to enhance the evaporation efficiency. Instead of traditional smearing method, the chemical wrapping method was used to better adhere the graphene nanoparticles on three different wick materials, which are utilized to absorb the solar energy and heat up the water on it to evaporate. The energy efficiencies were compared for the three different wick materials with/without wrapped graphene nanoparticles by two different reduction methods from graphene oxide (GO) solution with different concentrations. And the mass reduction rate, instantaneous energy efficiency and the absorptivity of different cases were measured and analyzed. Additionally, the morphological results of SEM images were given to furtherly support our analyses and discussions.

## 2. Experiment setup and Method

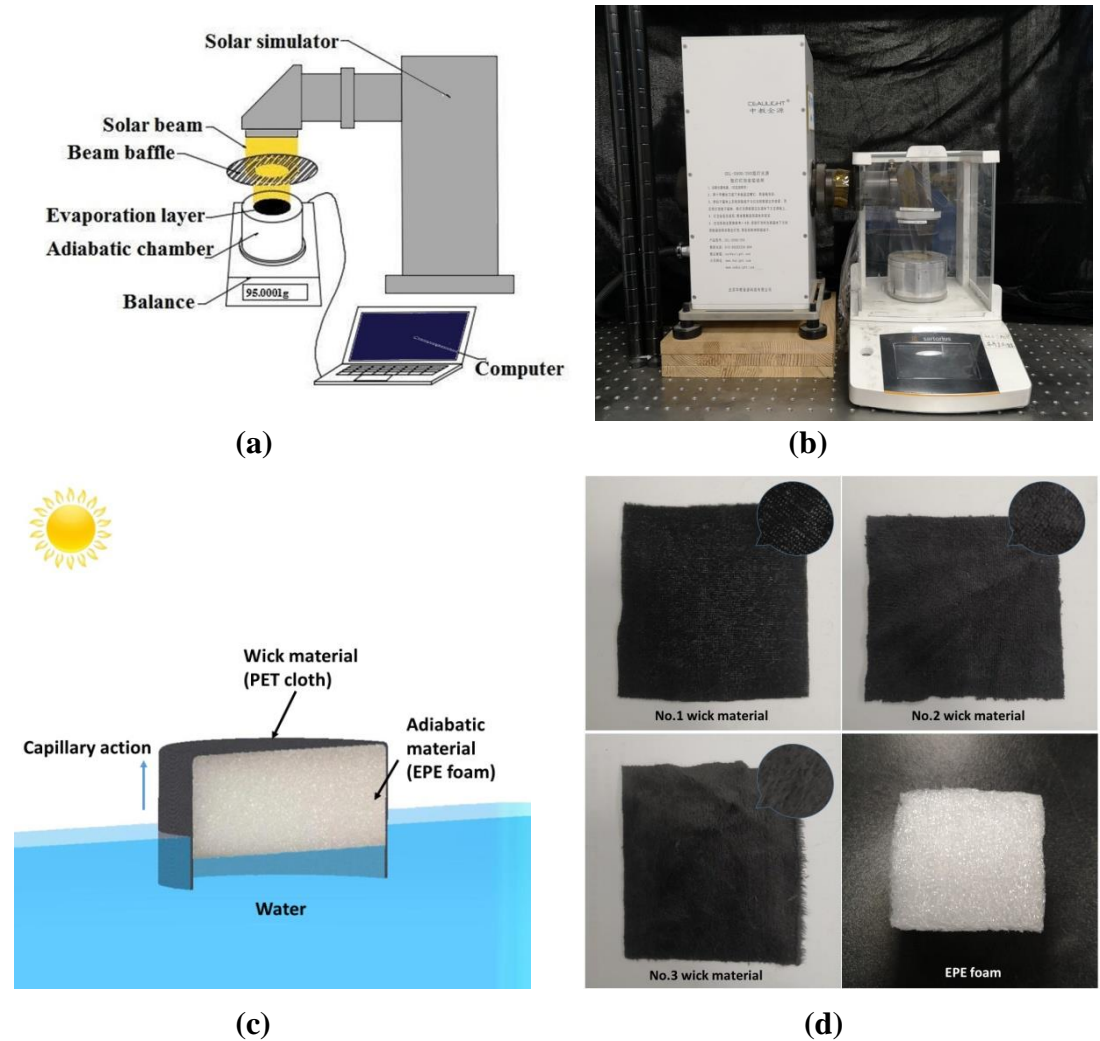

(a)      (b)

(c)      (d)

**Fig. 1** (a) Schematic diagram of experiment setup (b) Photograph of experiment setup (c) Schematic diagram of wick evaporation (d) Three wick materials and one adiabatic material

The schematic diagram and photograph of experiment setup are especially shown in Fig.1 (a) and (b). A solar simulator (CEL-S500, AM1.5) was used to generate the solar beam, the solar intensity was measured by a power meter (CEL-NP2000) and adjusted to 1kw/m$^2$. During the experiment, the edge of wick layer and the under lying bulk water were insulated by polyurethane foam (PU, 2.5cm thick) to minimize the heat loss. The diameter of the evaporation layer is 39mm, and a beam shelter slice was placed between the solar output window and the evaporation layer to cut the solar beam to the same diameter. The mass change during the evaporation process was measured by an electricity balance (Sartorius Practum 224), and the data were recorded by a computer via a USB cable. The room temperature and humidity during the experiment were controlled at 24 °C and 40% respectively.

Fig.1(c) shows the schematic diagram of wick evaporation. Bulk water is at the bottom of the entire structure. Capillary action wicks water to the top of the structure where the water is heated and evaporates. An adiabatic foam lies under the wick material to decrease the heat transfer between the bulk water and the heated water layer hence decreases the heat loss. Many daily materials can be chosen as the wick material such as cloth, tissue, paper and so on as long as the capillary action is strong enough to compensate the evaporation loss. The adiabatic material should be waterproof and have low density to float on the water.

Here, three different polyethylene terephthalate (PET) clothes were chosen as the wick materials, and the expanded polyethylene (EPE) foam was chosen as the adiabatic material as shown in Fig.1(d). Both the No.1 and No.3 wick materials are flannel clothes with many fine hairs , but the fine hairs of the No.3 cloth are denser than the No.1 cloth. The No.2 wick material is a kind of toweling cloth which has almost same density with the No.1 wick material but without fine hair. The three wick materials have good capillary action, and their price is quite cheap (~ $2/m$^2$). As a

adiabatic material, the EPE foam is suitable for our experiment system because it is light, adiabatic, cheap (~$10 m3), recyclable, anticorrosion and nontoxic. And it is safer compared with other common adiabatic material like rubber insulation cotton, expanded polystyrene (EPS).

The design of our experiment system is helpful to decrease the heat loss through the heat localization effect[19]. The use of adiabatic material leads to the thermal insulation between the evaporation region and the bulk water. Only the water in the thin wick material is heated, while the other water below the adiabatic material will not absorb heat. Thus almost all the heat is localized in the thin wick material, which results in high surface temperature and fast evaporation.

In order to further improve the evaporation efficiency, graphenen nanoparticles are added to the wick materials by chemical wrapping method instead of traditional smearing method [3]. Here, two methods were used to wrap graphene nanoparticles on the wick materials, including HI reduction method and microwave reduction method. The wick material was first soaked in graphene oxide (GO) solution at the concentration of 0.5 or 1 mg/ml for 5 seconds. Later, the wick material was put in vacuum drying oven at the temperature of 60 °C for 1 hour. For the HI reduction method, GO was reduced to graphene by soaking the GO-wrapped wick material in the HI solution at the temperature of 90 °C for 5 seconds. And then the wick material wrapped with reduced graphene oxide (rGO) was cleaned by deionized water for 2 times. Lastly, the wick material with rGO was put in vacuum drying oven at the temperature of 200 °C for 2 hours. For the microwave reduction method, the dried wick material with GO was put into the microwave oven to be reduced for 15~20 seconds. And then the wick material with rGO was cooled for a few seconds[32]. Thus the graphene nanoparticles were uniformly wrapped on the wick material. In our experiments, 16 different cases were carried out for solar steam generation as listed in Table 1.

**Table 1 Different experimental cases for solar steam generation**

| Name of cases | Detailed description |
|---|---|
| 1w | Pure No.1 wick material |
| 1w&0.5micro | No.1 wick material with wrapped graphene nanoparticles from 0.5 mg/ml GO solution by microwave reduction |
| 1w&1.0micro | No.1 wick material with wrapped graphene nanoparticles from 1.0 mg/ml GO solution by microwave reduction |
| 1w&0.5HI | No.1 wick material with wrapped graphene nanoparticles from 0.5 mg/ml GO solution by HI reduction |
| 1w&1.0HI | No.1 wick material with wrapped graphene nanoparticles from 1.0 mg/ml GO solution by HI reduction |
| 2w | Pure No.2 wick material |
| 2w&0.5micro | No.2 wick material with wrapped graphene nanoparticles from 0.5 mg/ml GO solution by microwave reduction |
| 2w&1.0micro | No.2 wick material with wrapped graphene nanoparticles from 1.0 mg/ml GO solution by microwave reduction |
| 2w&0.5HI | No.2 wick material with wrapped graphene nanoparticles from 0.5 mg/ml GO solution by HI reduction |
| 2w&1.0HI | No.2 wick material with wrapped graphene nanoparticles from 1.0 mg/ml GO solution by HI reduction |
| 3w | Pure No.3 wick material |
| 3w&0.5micro | No.3 wick material with wrapped graphene nanoparticles from 0.5 mg/ml GO solution by microwave reduction |
| 3w&1.0micro | No.3 wick material with wrapped graphene nanoparticles from 1.0 mg/ml GO solution by microwave reduction |
| 3w&0.5HI | No.3 wick material with wrapped graphene nanoparticles from 0.5 mg/ml GO solution by HI reduction |
| 3w&1.0HI | No.3 wick material with wrapped graphene nanoparticles from 1.0 mg/ml GO solution by HI reduction |
| 3w&1.0smear | No.3 wick material with graphite particles of 1.0mg/ml concentration by direct smearing method |

## 3. Results and Discussions

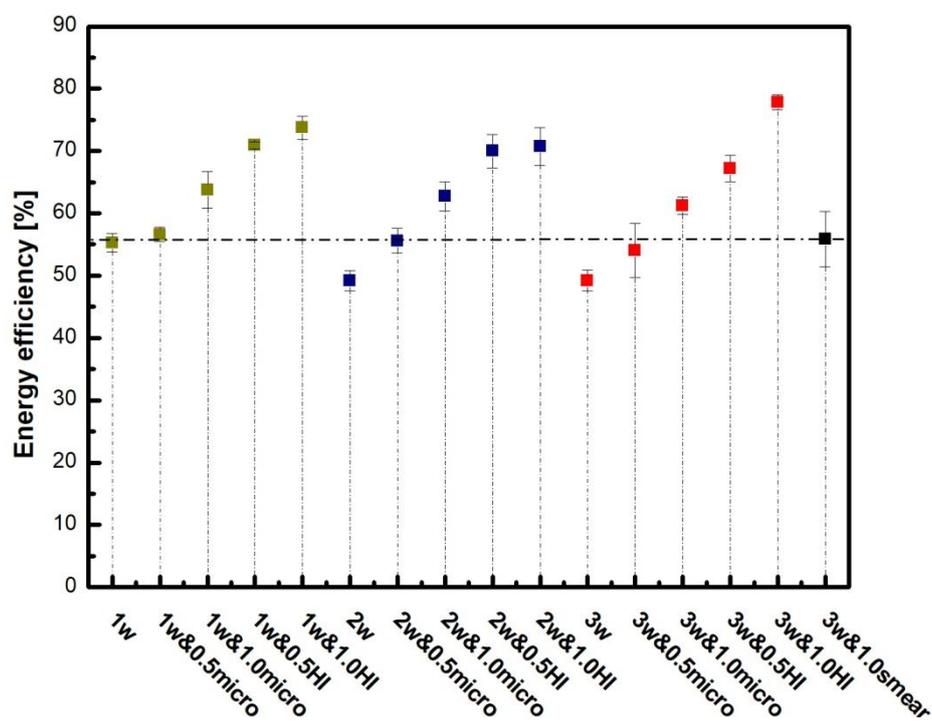

**Fig. 2** Comparison of the energy efficiencies of three different wick materials with/without wrapped graphene nanoparticles for the solar steam generator. Two different reduction methods, microwave reduction and HI reduction, are respectively used to wrap graphene nanoparticles by reducing graphene oxide solutions at two different concentrations, 0.5mg/ml or 1.0mg/ml. And the illustration of different experimental cases are listed in Table 1.

Fig.2 shows the comparison of the energy efficiencies of three different wick materials with/without wrapped graphene nanoparticles for the solar steam generation. As a typical representative, the different cases of No.3 wick material were firstly compared and analyzed. For pure No.3 wick material without nanoparticles (3w), the energy efficiency is around 47%. When the graphite particles of 1mg/ml concentration are added to the No.3 wick material by traditional smearing method (3w&1.0smear), the energy efficiency is around 56%. However, when graphene nanoparticles are added to the No.3 wick material by chemical wrapping method, the energy efficiency can be obviously enhanced. Especially for the No.3 wick material with wrapped graphene nanoparticles from 1mg/ml GO solution by HI reduction (3w&1.0HI), the energy efficiency can reach almost 80% which has increased 33% compared to the pure wick material without nanoparticles. Through the chemical wrapping method, the graphene nanoparticles can be evenly spread across the fibres of the wick material, which have better dispersity and stability than the way by traditional smearing method. Due to high thermal conductivity and large specific surface area, graphene nanoparticles have better solar absorption, heat localization and heat transfer. Therefore, the wick materials with wrapped graphene nanoparticles can enhance the energy efficiency dramatically.

For three different wick materials, due to their different configurations and properties, they have different influences on the energy efficiency. Both the No.1 and No.3 wick materials are flannel clothes, but the No.3 wick material has denser fine hairs. For the pure wick materials without particles, the energy efficiency of the No.3 wick material is lower than the No.1 wick material. Because the denser fine hairs result in thicker evaporation layer, which could bring more heat loss. However, after the graphene nanoparticles are wrapped on the wick materials, the energy efficiency of the No.3 wick material is nearly the same as that of the No.1 wick material under 0.5mg/ml concentration, and the efficiency of the No.3 wick material is even higher than that of the No.1 wick material under 1.0mg/ml concentration. The reason is that the denser fine hairs are conducive to wrap more graphene nanoparticles and thus contribute to the improvement of evaporation efficiency. The No.2 wick material is a kind of toweling cloth without fine hair. It can be seen that the energy efficiency of No.2 wick material is lower than that of the No.1 wick material under the same conditions. The reason is that the wick material with fine hair can provide three-dimensional contact area for wrapping nanoparticles and thus facilitate the water evaporation.

In addition, different reduction methods and different solution concentrations have some influence on the evaporation efficiency. For the same wick material, the efficiency by HI reduction method is better than that by microwave reduction method, and the explanation will be given in the subsequent analysis. For the same reduction method, the greater GO solution concentration can produce higher energy efficiency because more graphene nanoparticles can be wrapped on the wick materials.

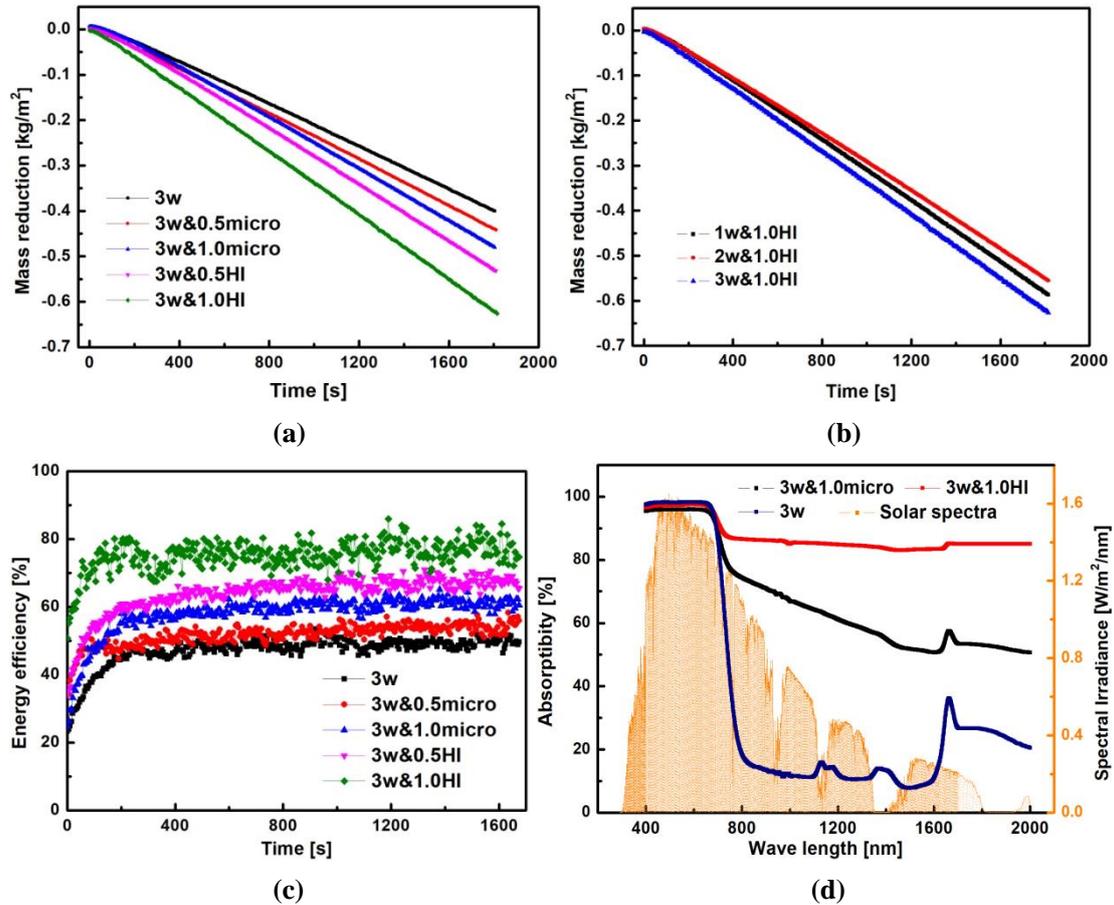

**Fig.3.** (a) The mass reduction due to evaporation versus time for different cases of the No.3 wick material: pure wick material (3w), wick material with wrapped nanoparticles from 0.5 or 1.0 mg/ml GO solution by microwave reduction or HI reduction (3w&0.5micro, 3w&1.0micro, 3w&0.5HI, 3w&1.0HI). (b) The mass reduction due to evaporation versus time for three different wick materials with wrapped nanoparticles from 1.0 mg/ml GO solution by HI reduction (1w&1.0HI, 2w&1.0HI, 3w&1.0HI). (c) The instantaneous energy efficiency for different cases of the No.3 wick material. (d) The absorptivity of pure No.3 wick material (3w) and the No.3 wick material with wrapped nanoparticles from 1.0 mg/ml GO solution by microwave reduction or HI reduction (3w&1.0micro, 3w&1.0HI).

Fig. 3(a) shows the comparison of the mass reduction due to evaporation versus time for different cases of the No.3 wick material. The mass reduction versus time represents the evaporation rate. Through the comparison, the evaporation rate of pure

No.3 wick material without nanoparticles is lowest, and the accumulated mass reduction due to the evaporation is less than 0.45g in half an hour. After the graphene nanoparticles are wrapped on the wick material, the evaporation rate is obviously enhanced. Especially for the No.3 wick material with wrapped nanoparticles from 1.0 mg/ml GO solution by HI reduction (3w&1.0HI), the accumulated mass reduction due to the evaporation reaches to 0.7g in half an hour, which is 1.6 times as that of pure No.3 wick material without nanoparticles. By comparing the two different reduction method, it is shown that the effect of wrapping nanoparticles by HI reduction is better than that by microwave reduction. And the evaporation rate of No.3 wick material with wrapped nanoparticles from 0.5mg/ml solution by HI reduction (3w&0.5HI) is even higher than that of wick material with wrapped nanoparticles from 1.0mg/ml solution by microwave reduction (3w&1.0micro).

Fig. 3(b) shows the comparison of mass reduction due to evaporation versus time for three different wick materials with wrapped nanoparticles from 1.0 mg/ml GO solution by HI reduction. After wrapping nanoparticles under the same conditions, the No.3 wick material can produce the highest evaporation rate among the three materials. Because the No.3 wick material has dense fine fairs, which can provide three-dimensional contact area for wrapping graphene nanoparticles and thus contribute to better evaporation.

The instantaneous energy efficiency of evaporation for different cases of the No.3 wick material is shown in Fig. 3(c). As we can see, the evaporation efficiency of pure No.3 wick material is gradually increasing with the measuring time and finally reaches to around 47%. For the cases of No.3 wick material with wrapped graphene nanoparticles, the evaporation efficiencies are faster to converge and get the maximum values. Especially for the No.3 wick material with wrapped nanoparticles from 1.0mg/ml solution by HI reduction(3w&1.0HI), the energy efficiency reaches stable during 5 minutes, and the stable efficiency is up to near 80%. Obviously, the utilization of wick material with wrapped nanoparticles is a very effective way to enhance the evaporation efficiency.

In order to better understand the improvement of evaporation efficiency, the absorptivity of three cases of No.3 wick material was measured. As shown in Fig. 3(d), the absorptivity of the three cases is almost the same and reaches to about 96% during the visible light band (400nm~760nm). Because the three cases are all black materials, which absorb visible light very well. However for the infrared light band (> 760 nm), the absorptivity of the three cases is obviously different due to their different compositions and microstructures. After the graphene nanoparticles are wrapped on the wick material, the absorptivity is obviously enhanced. And the absorptivity of the case by HI reduction is higher than that by microwave reduction. It is worth noting that, the solar spectrum spans both the visible light band and the infrared light band as shown by the orange line in Fig.3(d). Therefore, the difference of absorptivity is a comprehensive effect in both the visible light band and infrared light band. For the water evaporation, the higher absorptivity leads to higher energy efficiency.

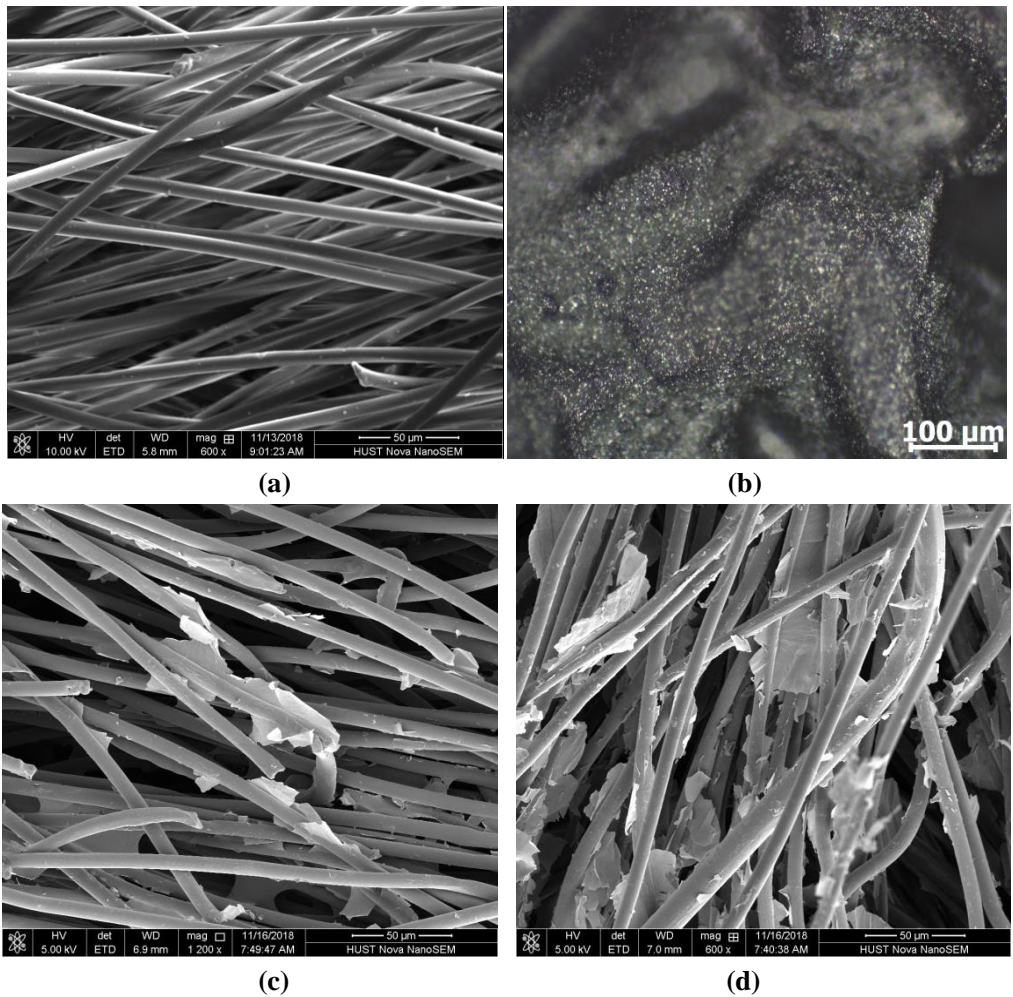

(a)　　　　　　　　　　　　　　(b)

(c)　　　　　　　　　　　　　　(d)

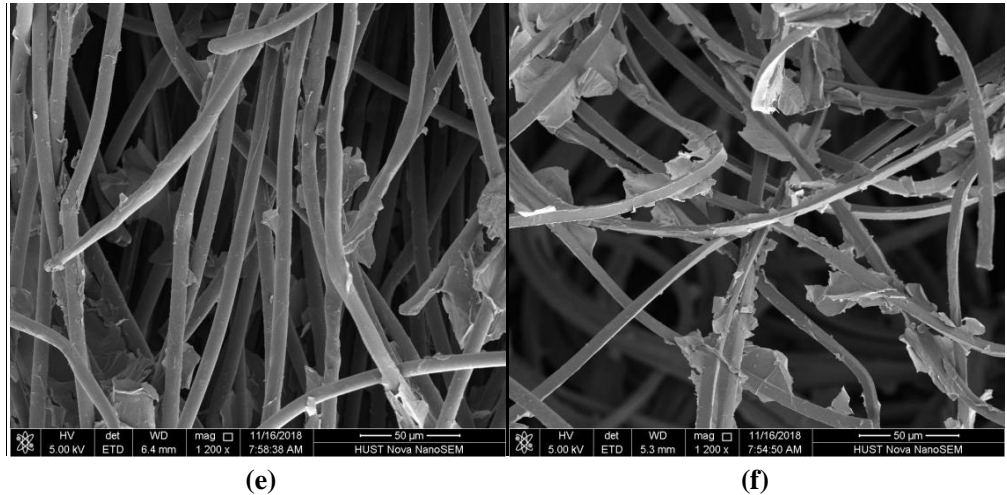

(e)                               (f)

**Fig.4.** (a) The SEM image of pure No.3 wick material (3w). (b) The optical microscopic image of the No.3 wick material covered by graphite particles through direct smearing method. (c) The SEM image of No.3 wick material with wrapped nanoparticles from 1.0mg/ml GO solution by HI reduction (3w&1.0HI). (d) The SEM image of No.3 wick material with wrapped nanoparticles from 1.0mg/ml GO solution by microwave reduction (3w&1.0micro). (e) The SEM image of No.1 wick material with wrapped nanoparticles from 1.0mg/ml solution by HI reduction (1w&1.0HI). (f) The SEM image of No.2 wick material with wrapped nanoparticles from 1.0mg/ml solution by microwave reduction (2w&1.0micro).

To intuitively exhibit the micro-scale morphology of wick materials with/without nanoparticles, the Scanning Electron Microscope (SEM) images of different materials are given as shown in Fig.4. We can see that the pure No.3 wick material is consist of many thin fibres, as shown in Fig.4(a). After the graphite particles are directly smeared on the No.3 wick material, the morphology of the material is exhibited in Fig.4(b). It is shown that the graphite particles are unevenly adhered to the wick material, and the particles tend to aggregate and deposit.

Fig.4(c) shows the SEM image of No.3 wick material with wrapped graphene nanoparticles from 1.0mg/ml GO solution by HI reduction. It can be seen that the graphene nanoparticles are uniformly wrapped across the fibres of wick material. The results indicate that the chemical wrapping method is able to make the nanoparticles more dispersive and stable than the way by direct smearing method. Fig.4(d) gives the

SEM image of No.3 wick material with wrapped graphene nanoparticles from 1.0mg/ml GO solution by microwave reduction. Although the graphene nanoparticles are also wrapped across the fibres of wick material, the wrapped nanoparticles by microwave reduction tend to gather together and form some sheet clusters, which are not conducive to heat transfer. Therefore, the energy efficiency by HI reduction method is better than that by microwave reduction method.

Fig.4(e) and Fig.4(f) respectively show the SEM images of No.1 wick material with wrapped nanoparticles by HI reduction and No.2 wick material with wrapped nanoparticles by microwave reduction. Compared with the corresponding cases of No.3 wick material, the No.1 wick material has cylindrical fibres of the same shape but fewer in quantity, while the No.2 wick material has prismatic fibres of different shape and fewer in quantity. Since less thin fibres have smaller contact area for wrapping graphene nanoparticles, the energy efficiency of No.1 wick material is lower than that of the No.3 wick material under the same conditions. In addition, because the wrapping effect of prismatic fibres is not good as that of cylindrical fibres, the energy efficiency of No.2 wick material is lower than that of the No.1 and No.3 wick materials. These morphological results furtherly confirm the previous analyses.

## 4. Conclusion

In conclusion, the energy efficiency of the proposed solar steam generator reaches up to 80% by using the wick material with wrapped graphene nanoparticles, which has increased 33% compared to the pure wick material without nanoparticles. Through the chemical wrapping method, the graphene nanoparticles can be evenly spread across the fibres of the wick material, which have better dispersity and stability than the way by traditional smearing method. Thus the energy efficiency can be dramatically enhanced by the wick materials with wrapped graphene nanoparticles due to better solar absorption, heat localization and heat transfer. Among the three wick materials, the No.3 wick material which is a kind of flannel clothe with dense fine hairs, is able to obtain the highest evaporation efficiency after wrapping graphene nanoparticles. The dense fine hairs are helpful to provide three-dimensional contact area for wrapping graphene nanoparticles and thus contribute to better evaporation. For the two different reduction methods, the results show that the energy efficiency by HI reduction method is better than that by microwave reduction method. In addition, the energy efficiency can be furtherly enhanced by increasing the concentration of graphene oxide solution used for obtaining graphene nanoparticles.

Our work indicates that different ways to use nanoparticles in solar steam generator may have different effects on energy efficiency, which is helpful to find an optimal method for using nanotechnology in practical application for harvesting solar energy. Also, the proposed solar steam generator by wick material with wrapped nanoparticles is a low-cost high-efficiency experiment system, which may have broad applications, such as desalination, water purification, power generation and so on.

## Acknowledgement

The work was supported by the National Natural Science Foundation of China No. 51606072 (C. D.), No. 51576076 (N. Y.) , No. 51711540031 (N. Y.), and Natural Science Foundation of Hubei Province No. 2017CFA046 (N. Y.).